\documentclass[12pt]{elsarticle}

\usepackage{amssymb}
\journal{European Journal of Operational Research}
\usepackage{amssymb}

\usepackage{rotating}

\usepackage{natbib}

\usepackage[tbtags]{amsmath}

\usepackage{eurosym}

\usepackage{multirow}

\usepackage{color}

\newcommand{\bfg}[1]{\mbox{\boldmath $#1$\unboldmath}}

\newcommand{\fraca}[2]{\displaystyle\frac{#1}{#2}}

\newcommand{\mm}[3]{\renewcommand{\arraystretch}{0.8}\begin{array}[t]{c}\mbox{#1}
\\ #2\end{array}\begin{array}[t]{c}#3\end{array}
\renewcommand{\arraystretch}{1}}

\def \R {{\rm I\kern -2.2pt R\hskip 1pt}}

\newenvironment{algorithm}[1]{\begin{algorithm1}{\bf (#1).}\em
$\;$ }{~\qed \end{algorithm1}}

\newtheorem{algorithm1}{Algorithm}[section]

\def\boxforqed{\rule{0.5em}{1.5ex}}
\def\qed{\ifmmode\squareforqed\else{\unskip\nobreak\hfil
        \penalty50\hskip1em\null\nobreak\hfil\boxforqed
         \parfillskip=0pt\finalhyphendemerits=0\endgraf}\fi}

\begin{document}

\begin{frontmatter}



\title{Robust Transmission Network Expansion Planning in Energy Systems: Improving Computational Performance}


\author[label1]{R.~M\'{\i}nguez\corref{cor1}}
\address[label1]{Independent consultant, C/Honduras 1, 13160, Torralba de Calatrava, Ciudad Real, Spain}
\author[label2]{R.~Garc\'{\i}a-Bertrand}
\address[label2]{Department of Electrical Engineering, University of Castilla-La Mancha,
Ciudad Real, Spain}
\cortext[cor1]{Corresponding author: rominsol@gmail.com, tlfn.: 00 34 926810046}

\begin{abstract}
% Text of abstract
In recent advances in solving the problem of transmission network expansion planning,   the use of robust optimization techniques has been put forward, as an alternative to stochastic mathematical programming methods, to make the problem tractable in realistic systems. Different sources of uncertainty have been considered, mainly related to the capacity and availability of generation facilities and demand, and making use of adaptive robust optimization models. The mathematical formulations for these models give rise to three-level mixed-integer optimization problems, which are solved using different strategies. Although it is true that these robust methods are more efficient than their stochastic counterparts, it is also correct that solution times for mixed-integer linear programming problems increase exponentially with respect to the size of the problem. Because of this, practitioners and system operators need to use computationally efficient methods when solving this type of problem. In this paper the issue of improving computational performance by taking different features from existing algorithms is addressed. In particular, we replace the lower-level problem with a dual one, and solve the resulting bi-level problem using a primal cutting plane algorithm within a decomposition scheme. By using this alternative and simple approach, the computing time for solving transmission expansion planning problems has been reduced drastically. Numerical results in an illustrative example, the IEEE-24 and IEEE 118-bus test systems demonstrate that the algorithm is superior in terms of computational performance with respect to existing methods.
%
%A comprehensive case study based on the self-scheduling problem of
%a power producer is used to illustrate the capabilities of the
%proposed  methodology. Appropriate conclusions are finally drawn.
\end{abstract}

%%%\begin{keywords}
%%%% keywords here, in the form: keyword \sep keyword
%%%Stochastic programming \sep Conic programming
%%%and interior point methods \sep Decision analysis under uncertainty \sep Reliability analysis \sep Robust optimization
%%%% PACS codes here, in the form: \PACS code \sep code
%%%
%%%%% MSC codes here, in the form: \MSC code \sep code
%%%%% or \MSC[2008] code \sep code (2000 is the default)
%%%
%%%\end{keywords}
\end{frontmatter}

% main text

%
\section{Introduction}
The aim of Transmission Network Expansion Planning (TNEP) is to resolve the issue of how to expand or enhance an existing electricity transmission network in order to adequately satisfy demand for a given horizon. The main difficulty of this problem is  decision-taking with the great amount of uncertainty associated with i) demand, ii) generation with renewable energy sources, such as wind and solar power plants, and iii) equipment failure. Two main frameworks have been used to tackle these uncertainties, stochastic programming \citep{BirgeL:97} and robust optimization \citep{Soyster:73,El-GhaouiL:97,El-GhaouiOL:98,Ben-TalN:98,Ben-TalN:99,Ben-TalN:00,BertsimasS:04}.

In stochastic programming problems, uncertain data is assumed to follow a given probability distribution and is usually dealt with by using scenarios or finite sampling from the joint probability density function. Examples of the successful application of stochastic programming in TNEP problems are given by \citep{CarrionAA:07,LopezPQ:07,GarcesCGR:09,RohSW:09} among others. However, the number of scenarios needed to represent the actual stochastic processes can be very large, which may result in intractable problems. Conversely, robust optimization can avoid the intractability issue related to stochastic programming approaches. Note that tractability is not the only difficulty stemming from stochastic programming, but it is the most important aspect from the point of view given in this paper. Examples of the successful application of robust optimization in TNEP problems are given by \citep{WuCX:08,YuCW:11}. However, in this paper we focus on the seminal works proposed by \citep{Jabr:13}, \citep{ChenWWHW:14} and \citep{RuizC:14}.

The three works on which our proposal is based formulate the problem using an Adaptive Robust Optimization (ARO) framework \citep{ThieleTE:10,BertsimasLSZZ:13}. Therefore, all of them use three-level formulations: i) the first level minimizes the cost of expansion (\citep{ChenWWHW:14} also minimizes the maximum regret), the decision variables for this level are those associated with construction or expansion of lines, ii) the second level selects the realization of the uncertain parameters that maximizes the system's operating costs within the uncertainty set, the variables related to this level are the uncertain parameters, i.e., generation capacity and demand, and iii) in the third level the system operator selects the optimal decision variables to minimize operating costs for given values of first and second level variables. Note that in this paper we do not consider contingencies explicitly, although, they could be taken into account within  robust formulation following the guidelines set out by \citep{StreetOA:11}.

Jabr \citep{Jabr:13} merges the second and third levels into one single-level maximization problem (so called subproblem in this paper) by using the dual of the third level. It takes advantage of the fact that all uncertain parameters need to be equal to their upper or lower limits in the worst case scenario, which allows the use of binary variables to deal with these uncertain parameters. The limitation of this simplification is that the uncertainty budget, typically used in robust optimization, must be an integer. However, in our opinion, this does not detract from the benefits of robust optimization ,while it does simplify the resolution of the problem substantially. In contrast, in references \citep{ChenWWHW:14,RuizC:14}  the second and third levels are merged into one single level maximization problem by using the Karush-Kuhn-Tucker (KKT) conditions for the third level. In this particular case, the number of constraints, continuous and binary variables of the subproblem increase with respect to the other approach.

Regarding the resolution of the resulting bi-level problem, Jabr \citep{Jabr:13} sets out a Benders \citep{ConejoCMG:06} approach where the dual information from subproblems is used to make additional Benders cuts. The main drawback of this approach is the slow convergence typical of this type of decomposition algorithms, which forced the author to include additional linear constraints in order to improve convergence. On the other hand, Cheng et al. and Ruiz and Conejo \citep{ChenWWHW:14,RuizC:14} applied a constraint-and-column \citep{ZengZ:13} generation method solely based on primal cuts. This method is computationally advantageous with respect to  Benders decomposition and converges in a small number of iterations (authors reported solution results with only 3-4 iterations required). Note that the methodologies proposed by \citep{ChenWWHW:14} and \cite{RuizC:14} are basically the same from a mathematical point of view, although, they analyse different situations and Ruiz and Conejo \cite{RuizC:14} set out a different way to construct the uncertainty sets.

In this paper an alternative approach to solving the three-level mixed-integer optimization problem associated with TNEP is set out. The concept elaborated on in this paper is to combine the way the inner level problems (subproblems) are solved by \citep{Jabr:13} and the constraint-and-column generation method used by \citep{ChenWWHW:14} and \cite{RuizC:14} for the first level problem (master problem). Note that although the master problem solved by means of primal cuts is very efficient, computing times associated with the approaches adopted by \citep{ChenWWHW:14} and \cite{RuizC:14}  increase exponentially with respect to the size of the problem. This is due to the subproblem, which consumes most of the resolution time due to the use of the KKT optimality conditions for the third level in the second level problem. The resulting model  simultaneously calculates the third level primal and dual variables, and as the primal variables are not required to construct the primal cuts of the master, this alternative consisting in use of the KKT conditions is inefficient.

The remainder of the paper is structured as follows. Section~\ref{s1} describes the adaptive robust formulation of the TNEP problem in compact matrix form, and discusses the approaches for finding solutions  set out in references \citep{Jabr:13}, \citep{ChenWWHW:14} and \cite{RuizC:14}. The detailed formulation is only given in the \ref{detfor}. In Section~\ref{s2} the definition of the uncertainty set and the description of the approach for finding a solution are set out. Numerical results for different networks are given in Section~\ref{CaseStudy} and compared with those obtained using the method proposed by \citep{ChenWWHW:14} and \cite{RuizC:14}. Finally, the paper is concluded in Section~\ref{Conclusions}.

\section{Transmission network expansion planning problem: ARO compact formulation}\label{s1}
%
%Note that for clarity in the exposition, we adopt the same nomenclature used by \citep{RuizC:14}. Thus,
According to references \citep{Jabr:13}, \citep{ChenWWHW:14} and \cite{RuizC:14}, the robust TNEP problem can be written in compact matrix form as the following three-level mathematical programming problem:
\begin{equation}\label{eq1}
  \mm{Minimize}{{\bfg x}}{\left({\bfg c}^T{\bfg x}+\mm{Maximum}{{\bfg d}\in {\bfg D}}{\mm{Minimum}{{\bfg y}\in \Omega({\bfg x},{\bfg d})}{{\bfg b}^T}{\bfg y}}\right)}
\end{equation}
subject to
\begin{eqnarray}
  {\bfg c}^T{\bfg x} & \le & \Pi \label{eq2}\\
  {\bfg x} & \in & \{0,1\}, \label{eq3}
\end{eqnarray}
where ${\bfg x}$ is the vector of first stage binary variables representing the {\em investment} vs {\em no investment} in reinforcing or building new lines, ${\bfg c}$ is the investment cost vector, ${\bfg d}$ is the  second stage continuous variables vector representing the random or uncertain parameters, i.e. generation capacities and level of demand, ${\bfg D}$ is the uncertainty set, ${\bfg b}$ is the vector that includes operating costs, and ${\bfg y}$ is the  third stage continuous variables vector referring to operating variables. These operating variables include power consumed, power flows, power produced by generating units, load shed by demand and voltage angles at buses (see \ref{detfor} for a detailed description of the formulation). $\Pi$ represents the maximum budget for investment in transmission expansion. Finally, $\Omega({\bfg x},{\bfg d})$ defines the feasibility region for the operating variables ${\bfg y}$, as a function of investment decisions ${\bfg x}$ and given realizations of the uncertain parameters ${\bfg d}$, as follows:
\begin{equation}\label{eq4}
  \Omega({\bfg x},{\bfg d})=\left\{
  \begin{array}{rcl}
    {\bfg A}{\bfg x}+{\bfg B}{\bfg y} & = & {\bfg E}:{\bfg \lambda} \\
    {\bfg F}{\bfg x}+{\bfg G}{\bfg y} & \le & {\bfg K}:{\bfg \mu}\\
    {\bfg I}_{\rm eq} {\bfg y} & = & {\bfg d}:{\bfg \alpha}\\
     {\bfg I}_{\rm ineq} {\bfg y} & \le & {\bfg d}:{\bfg \varphi},
  \end{array}
  \right.
\end{equation}
where ${\bfg A}$, ${\bfg B}$, ${\bfg E}$, ${\bfg F}$, ${\bfg G}$ and ${\bfg K}$ are matrices with constant parameters dependent on the network configuration and element characteristics, ${\bfg I}_{\rm eq}$ selects the components of ${\bfg y}$ that are equal to the uncertain parameters (demand), and ${\bfg I}_{\rm ineq}$ selects the components of ${\bfg y}$ that are limited by the uncertain parameters (i.e. maximum power generation and maximum load shedding). The first set of equality constraints correspond to those enforcing the power balance at every bus, the power flow through each line, and fixing the voltage angle of the reference bus. The second set of inequality constraints  are associated with line flow limits, and limits on the voltage angles at every bus. Note that ${\bfg \lambda}$, ${\bfg \mu}$, ${\bfg \alpha}$ and ${\bfg \varphi}$ are the dual variable vectors associated with each set of constraints, respectively.

For a detailed physical interpretation of the mathematical formulation (\ref{eq1})-(\ref{eq3}) we recommend reference \cite{RuizC:14}.

\subsection{The bi-level approach put forward by Jabr}
Jabr \citep{Jabr:13} proposes tackling the problem (\ref{eq1})-(\ref{eq3}) by decomposing and iteratively solving a subproblem and a master problem. The master variables correspond to ${\bfg x}$, i.e. the  first stage binary variables vector . Thus, for given values of these master variables, the subproblem corresponds to:
\begin{equation}\label{subproblem}
  \mm{Maximum}{{\bfg d}\in {\bfg D}}{\mm{Minimum}{{\bfg y}\in \Omega({\bfg x},{\bfg d})}{{\bfg b}^T}{\bfg y}}.
\end{equation}

Considering that the dual problem associated with the third-level is equal to:
\begin{equation}\label{dual1}
  \mm{Maximize}{{\bfg \lambda},{\bfg \mu},{\bfg \alpha},{\bfg \varphi}}{({\bfg E}-{\bfg A}{\bfg x})^T{\bfg \lambda}-({\bfg K}-{\bfg F}{\bfg x})^T{\bfg \mu}+{\bfg d}^T({\bfg \alpha}-{\bfg \varphi})}
\end{equation}
subject to
\begin{eqnarray}
  {\bfg B}^T{\bfg \lambda}-{\bfg G}^T{\bfg \mu}+ {\bfg I}_{\rm eq}^T{\bfg \alpha}-{\bfg I}_{\rm ineq}^T{\bfg \varphi}& = & {\bfg b} \label{dual2}\\
 {\bfg \mu} & \ge & {\bf 0} \label{dual3}\\
  {\bfg \varphi} & \ge & {\bf 0} \label{dual4},
\end{eqnarray}
it can be substituted into (\ref{subproblem}), which results in the following single level maximization problem:
\begin{equation}\label{subproblem1}
  \mm{Maximize}{{\bfg d},{\bfg \lambda},{\bfg \mu},{\bfg \alpha},{\bfg \varphi}}{f^{\rm dual}=({\bfg E}-{\bfg A}{\bfg x})^T{\bfg \lambda}-({\bfg K}-{\bfg F}{\bfg x})^T{\bfg \mu}+{\bfg d}^T({\bfg \alpha}-{\bfg \varphi})}
\end{equation}
subject to (\ref{dual2}), (\ref{dual3}), (\ref{dual4}), and
\begin{equation}\label{subproblem2}
{\bfg d}\in {\bfg D}.
\end{equation}

Subproblem (\ref{subproblem1})-(\ref{subproblem2}) is a bilinear mathematical programming problem, which can be  linearized and transformed into a mixed-integer linear mathematical programming problem at the expense of introducing binary variables associated with the uncertainty set (see \citep{Jabr:13} for more details). This  simplification consists in  assuming that the uncertain parameters are either at the nominal, upper or lower limits of their uncertainty range, as shown by \citep{WuCX:08} and \citep{Jabr:13}.

The optimal solution for the subproblem (\ref{subproblem1})-(\ref{subproblem2}) provides the dual variable information and the uncertain parameter values to construct Benders cuts for the master problem. The main problem of this type of decomposition algorithm is the slow convergence property, which forced the author to include additional linear constraints obtained from a number of previously computed realizations of the uncertainty vector. These additional cuts accelerate convergence.

%%%, which corresponds to the following optimization problem at iteration $k$:
%%%%
%%%\begin{equation}\label{master1}
%%%  \mm{Minimize}{{\bfg x}}{{\bfg c}^T{\bfg x}+\gamma}
%%%\end{equation}
%%%subject to
%%%\begin{eqnarray}
%%%% \nonumber to remove numbering (before each equation)
%%%  \gamma \!&\! \ge \!&\! ({\bfg E}-{\bfg A}{\bfg x})^T{\bfg \lambda}^{(i)}-({\bfg K}-{\bfg F}{\bfg x})^T{\bfg \mu}^{(i)}+{\bfg d}_{(i)}^T({\bfg \alpha}^{(i)}-{\bfg \varphi}^{(i)});\;\forall i<k\label{master2}\\
%%%   {\bfg c}^T{\bfg x}\! &\! \le \!&\! \Pi \label{master3}\\
%%%  {\bfg x} \!&\! \in \!&\! \{0,1\}. \label{master4}
%%%\end{eqnarray}

\subsection{Bi-level approach proposed by Chen et al. and Ruiz and Conejo}
Chen et al. \citep{ChenWWHW:14} and Ruiz and Conejo \cite{RuizC:14} also propose dealing with problem (\ref{eq1})-(\ref{eq3}) by decomposing and iteratively solving a subproblem and a master problem. The master variables correspond to ${\bfg x}$, i.e. the vector of first stage binary variables. However, these authors set out a different solution strategy. In particular,  the third-level minimization problem is substituted into the subproblem (\ref{subproblem}) because of  its KKT conditions, with which the following single level maximization problem is obtained:
\begin{equation}\label{subproblem3}
  \mm{Maximize}{{\bfg d},{\bfg y},{\bfg \lambda},{\bfg \mu},{\bfg \alpha},{\bfg \varphi}}{{\bfg b}^T{\bfg y}}
\end{equation}
subject to
\begin{eqnarray}
 {\bfg A}{\bfg x}+{\bfg B}{\bfg y} & = & {\bfg E} \label{subproblem4}\\
  {\bfg I}_{\rm eq} {\bfg y} & = & {\bfg d}\label{subproblem5}\\
 0 & = & {\bfg b} +{\bfg B}^T{\bfg \lambda}-{\bfg G}^T{\bfg \mu}+ {\bfg I}_{\rm eq}^T{\bfg \alpha}-{\bfg I}_{\rm ineq}^T{\bfg \varphi} \label{subproblem6}\\
 0 & \le & {\bfg K}-{\bfg F}{\bfg x}-{\bfg G}{\bfg y} \perp  {\bfg \mu}  \ge  {\bf 0} \label{subproblem7} \\
 0 & \le & {\bfg d}- {\bfg I}_{\rm ineq} {\bfg y} \perp  {\bfg \varphi} \ge  {\bf 0}\label{subproblem8} \\
{\bfg d} & \in & {\bfg D},\label{subproblem9}
\end{eqnarray}
where constraint (\ref{subproblem6}) results from differentiating the Lagrangian of the third-level problem with respect to third-level variables ${\bfg y}$, and constraints  (\ref{subproblem7}) and (\ref{subproblem8}) represent the complementary conditions associated with inequality constraints from (\ref{eq4}).

Subproblem (\ref{subproblem3})-(\ref{subproblem9}) is a nonlinear mathematical programming problem, which can be linearized and transformed into a mixed-integer linear mathematical programming problem at the expense of using the Fortuny-Amat \citep{FortunyAmatM:81} transformation, which requires including a binary variable for each inequality constraint. For more details about Fortuny-Amat transformation related to this specific application, see reference \cite{RuizC:14}.

The optimal solution of subproblem (\ref{subproblem3})-(\ref{subproblem9}) provides the uncertain parameter values ${\bfg d}_{(k)}$ to construct primal cuts for the master problem at iteration $k$, which corresponds to the following optimization problem:
\begin{equation}\label{master5}
  \mm{Minimize}{{\bfg x},{\bfg y}_{(i)};\;\forall i= 1,\ldots,k-1}{{\bfg c}^T{\bfg x}+\gamma}
\end{equation}
subject to
\begin{eqnarray}
\gamma & \ge & {\bfg b}^T{\bfg y}_{(i)};\;\forall i= 1,\ldots,k-1\label{master7}\\
\gamma & \ge & 0\\
   {\bfg A}{\bfg x}+{\bfg B}{\bfg y}_{(i)} & = & {\bfg E};\;\forall i= 1,\ldots,k-1\label{master6}\\
    {\bfg F}{\bfg x}+{\bfg G}{\bfg y}_{(i)} & \le & {\bfg K};\;\forall i= 1,\ldots,k-1\label{master8}\\
    {\bfg I}_{\rm eq} {\bfg y}_{(i)} & = & {\bfg d}_{(i)};\;\forall i= 1,\ldots,k-1\label{master9}\\
     {\bfg I}_{\rm ineq} {\bfg y}_{(i)} & \le & {\bfg d}_{(i)};\;\forall i= 1,\ldots,k-1\label{master10}\\
   {\bfg c}^T{\bfg x} & \le & \Pi \label{master11}\\
  {\bfg x} & \in & \{0,1\}. \label{master12}
\end{eqnarray}

This master problem, besides variable $\gamma$, includes one operating variable vector ${\bfg y}_{(i)};\;\forall i= 1,\ldots,k-1$ for each realization of the uncertain parameters obtained from subproblem (\ref{subproblem3})-(\ref{subproblem9}). Unlike the Benders master problem set out in \cite{Jabr:13}, this algorithm converges in a few iterations without needing to include additional primal cuts besides those included by the iterative process itself. Note that the additional constraints included by Jabr to improve convergence of the Benders algorithm, obtained from a prespecified number of previously computed realizations of the uncertainty vector, is a similar strategy with respect to the column-and constrained generation method. However, in the latter case, those realizations are obtained by the process itself, which speeds up convergence.

\section{Proposed solution technique}\label{s2}
In this section the solution technique is set out and justified in detail. However, before focusing on the mathematical description of the solution method, we will justify the uncertainty set type used in this paper.

\subsection{Uncertainty set}
To deal with uncertainty we use cardinality constrained uncertainty sets \citep{BertsimasS:04}. In particular, we use the same definition of uncertainty set as that given in \citep{Jabr:13} but distributed by type of random parameter (generation capacity $G$ or demand $D$) and per region $r$:
\begin{equation}
  {\bfg D} = \left\{
  \begin{array}{ccc}
    d^G_{i} & = & \bar d^G_{i}+\hat d^G_{i}z^G_{i};\;\forall i\in \Omega^{\rm G} \\
    d^D_{i} & = & \bar d^D_{i}+\hat d^D_{i}z^D_{i};\;\forall i\in \Omega^{\rm D}
  \end{array}
  \right.
\end{equation}
where $d^G_{i}$ is the uncertain generation limit related to the $i$th variable within vector ${\bfg d}$, $\bar d^G_{i}$ is the corresponding nominal value, $\hat d^G_{i}$ is the maximum positive distance from the nominal value that can take the random parameter, and $\Omega^{\rm G}$ is the set of indices of the generating units. Analogously, $d^D_{i}$, $\bar d^D_{i}$, $\hat d^D_{i}$ and $\Omega^{\rm D}$ correspond to the same values but for demand. Additionally, the following constraints associated with budget uncertainty must hold:
  \begin{eqnarray}
    \sum_{i\in (\Omega^{\rm G}\cap\; \Omega^{\rm r})} |z^G_{i}|  &\le &\Gamma_r^G;\;\forall r  \\
    \sum_{i\in (\Omega^{\rm D}\cap\; \Omega^{\rm r})} |z^D_{i}|  &\le &\Gamma_r^D;\;\forall r,
  \end{eqnarray}
where $\Omega^{\rm r}$ is the set of indices for region $r$, $z^G_{i}$ and $z^D_{i}$ are auxiliary continuous variables with the following characteristics $|z^G_{i}|\le 1;\forall i\in \Omega^{\rm G}$ and $|z^D_{i}|\le 1;\forall i\in \Omega^{\rm D}$, and $\Gamma_r^G$ and $\Gamma_r^D$ are the maximum number of random parameters for generation capacity and demand, respectively, which may reach their lower or upper limits within region $r$. We include the discrimination by region according to that set out by \citep{RuizC:14}.

It must be stressed that the auxiliary variables $z^G_{i}$ and $z^D_{i}$ are initially assumed to be continuous. However, in order to solve the robust TNEP such as by \citep{Jabr:13}, these variables will be considered as binary. The only limitation introduced by this simplification is that uncertainty budgets $\Gamma_r^G$ and $\Gamma_r^D$ must be integer values, which in our opinion does not detract from the benefits of robust optimization from a practical perspective.

In addition to the simplification proposed by Jabr \citep{Jabr:13}, we also consider the  discovery of Ruiz and Conejo \citep{RuizC:14} consisting of the worst outcome in ``nature'' is for generation capacities to be as low as possible whilst demand loads is as high as possible. According to this, our uncertainty sets are finally defined as follows:
 \begin{eqnarray}
    d^G_{i} & = & \bar d^G_{i}-\hat d^G_{i}z^G_{i};\;\forall i\in \Omega^{\rm G} \label{uncer1}\\
    d^D_{i} & = & \bar d^D_{i}+\hat d^D_{i}z^D_{i};\;\forall i\in \Omega^{\rm D} \label{uncer3}\\
     \sum_{i\in (\Omega^{\rm G}\cap\; \Omega^{\rm r})} z^G_{i}  &\le &\Gamma_r^G;\;\forall r  \\
    \sum_{i\in (\Omega^{\rm D}\cap\; \Omega^{\rm r})} z^D_{i}  &\le &\Gamma_r^D;\;\forall r\\
    z^G_{i}&\in &\{0,1\};\forall i\in \Omega^{\rm G}\\
    z^D_{i}&\in &\{0,1\};\forall i\in \Omega^{\rm D}.\label{uncer2}
  \end{eqnarray}

Alternatively, instead of uncertainty budgets associated with generation, demand and regions $\Gamma_r^G$ and $\Gamma_r^D$, a unique uncertainty for each region $\Gamma_r$, or for the system $\Gamma$ could be used instead.

Note that we do not use the uncertainty set defined by Ruiz and Conejo \citep{RuizC:14} because in its current form it does not provide appropriate interpretation with respect to robustness. The definition of the uncertainty set given by \citep{RuizC:14} is:
  \begin{eqnarray}
    \fraca{\sum_{i\in \Omega^{\rm r}} (d^{G^{\rm max}}_{i}-d^{G}_{i})}{\sum_{i\in \Omega^{\rm r}} (d^{G^{\rm max}}_{i}-d^{G^{\rm min}}_{i})} &\le &\Gamma_r^G;\;\forall r  \\
    \fraca{\sum_{i\in \Omega^{\rm r}} (d^{D}_{i}-d^{D^{\rm min}}_{i})}{\sum_{i\in \Omega^{\rm r}} (d^{D^{\rm max}}_{i}-d^{D^{\rm min}}_{i})} &\le &\Gamma_r^D;\;\forall r.
  \end{eqnarray}

This formulation presents the following interpretation problem. If $\Gamma_r^G=1$ or $\Gamma_r^D=1$, the uncertainty variables associated with generation take the minimum possible values $d^{G^{\rm min}}_{i}$ and those random variables related to load demand take the maximum possible values $d^{D^{\rm max}}_{i}$. This scenario corresponds to the maximum level of uncertainty \citep{Soyster:73} as pointed out by \citep{RuizC:14}. However, if $\Gamma_r^G=0$ or $\Gamma_r^D=0$, the uncertainty variables associated with generation take the maximum values possible $d^{G^{\rm max}}_{i}$ and those random variables related to load demand take the minimum possible values $d^{D^{\rm min}}_{i}$. This scenario does not correspond to the {\em no uncertainty} case as pointed out in \citep{RuizC:14}, but rather to the most favourable one from the system operation perspective, and capacity expansion planning using this uncertainty budget would probably result in null investments and the resulting network could be congested for all possible values of the random parameters within the uncertainty set. Because of this it is difficult to interpret intermediate situations $0<\Gamma_r^G<1$ and $0<\Gamma_r^D<1$.

\subsection{Proposed decomposition method}
Once the uncertainty set is properly defined, we focus on our methodological proposal for solving problem (\ref{eq1})-(\ref{eq3}). It is based on the following observations:
\begin{enumerate}
  \item The subproblem defined by Jabr \citep{Jabr:13} is much more efficient than that proposed by Cheng et al. \citep{ChenWWHW:14} and Ruiz and Coenjo \cite{RuizC:14}. Note that the number of binary variables is equal to the number of uncertain parameters (cardinality of $\Omega^G$ plus cardinality of $\Omega^D$, i.e. $n_g+n_d$), which are the only variables involved including the dual variables of the third-level problem and the uncertain parameters themselves. Conversely, the subproblem defined using the KKT conditions of the third level problem includes the uncertain parameters, the primal and dual variables of the third-level problem. The number of binary variables associated with the maximum generation capacities and upper bounds of load-shedding are just equal to the number of uncertain parameters, and we have to include two additional binary variables per limit associated with the following variables: line flow, voltage angles, power generation and load-shedding. This results in a much bigger and complex subproblem to solve.
  \item Furthermore, the only variables required for the master problem set out by \citep{ChenWWHW:14} and \cite{RuizC:14} are the uncertain parameters. Therefore,  calculation of the primal variables within the subproblem is useless and inefficient.
  \item The master problem proposed by \citep{ChenWWHW:14} and \cite{RuizC:14} based on primal cuts is at an advantage in computing terms  with respect to the Benders master problem based on dual cuts set out by \citep{Jabr:13}. It converges in a few iterations and does not require including additional cuts artificially.
\end{enumerate}

For the reasons given above, we use as a subproblem a slight modification of that proposed by Jabr \citep{Jabr:13} and as a master problem that set out by \citep{ChenWWHW:14} and \cite{RuizC:14}, as follows:
\begin{description}
  \item [Subproblem:] The subproblem is aimed at obtaining for given values of the investment variables, i.e. for a given network configuration, the worst outcome for uncertain parameters producing the maximum optimum operational cost. In terms of equations, it consists of the maximization of the objective function (\ref{subproblem1}) subject to constraints (\ref{dual2}), (\ref{dual3}), (\ref{dual4}) and the uncertainty set definition (\ref{uncer1})-(\ref{uncer2}).
  \item[Master problem:] Given the worst outcome for the uncertain parameters from subproblems, the master problem is aimed at selecting the investment, i.e. the network configuration, that minimizes both the investment and the least desirable (maximum) operational costs associated with each set of values of the uncertain parameters. In terms of equations, it consists of minimizing the objective function (\ref{master5}) subject to constraints (\ref{master7})-(\ref{master12}).
\end{description}

The only additional detail required to properly define the method is the linearization of the bilinear term included in the objective function of the subproblem, i.e. ${\bfg d}^T({\bfg \alpha}-{\bfg \varphi})$. Considering equations (\ref{uncer1})-(\ref{uncer3}), this bilinear term becomes:
\begin{equation}\label{lineali}
\begin{array}{rcl}
  {\bfg d}^T({\bfg \alpha}-{\bfg \varphi}) & = & \displaystyle\sum_{\forall i\in \Omega^G} d_i^G\varphi_i^{\rm G}+\displaystyle\sum_{\forall i\in \Omega^D} e_i d_i^D\varphi_i^{\rm D}+\displaystyle\sum_{\forall i\in \Omega^D} d_i^D\alpha_i^{\rm D}\\
  & = & \displaystyle\sum_{\forall i\in \Omega^G} (\bar d^G_{i}-\hat d^G_{i}z^G_{i})\varphi_i^{\rm G}+\displaystyle\sum_{\forall i\in \Omega^D} e_i(\bar d^D_{i}+\hat d^D_{i}z^D_{i})\varphi_i^{\rm D}\\
    & + & \displaystyle\sum_{\forall i\in \Omega^D} (\bar d^D_{i}+\hat d^D_{i}z^D_{i})\alpha_i^{\rm D}=\displaystyle\sum_{\forall i\in \Omega^G} (\bar d^G_{i}\varphi_i^{\rm G}-\hat d^G_{i}z^G_{i}\varphi_i^{\rm G})\\
    & +  &  \displaystyle \sum_{\forall i\in \Omega^D} e_i(\bar d^D_{i}\varphi_i^{\rm D}+\hat d^D_{i}z^D_{i}\varphi_i^{\rm D})+ \displaystyle\sum_{\forall i\in \Omega^D} (\bar d^D_{i}\alpha_i^{\rm D}+\hat d^D_{i}z^D_{i}\alpha_i^{\rm D}).
\end{array}
\end{equation}

Note that for questions of consistency, the negative sign on the compact expression for ${\bfg \varphi}$ is implicitly included in its components $\varphi_i^{\rm G}$ and $\varphi_i^{\rm D}$, which are defined as negative in the detailed formulation given in the \ref{detfor}. $e_i$ is the percentage of load shed by demand $i$.

The terms to be linearized are $z^G_{i}\varphi_i^{\rm G};\;\forall i\in \Omega^G$, $z^D_{i}\varphi_i^{\rm D};\;\forall i\in \Omega^D$, and $z^D_{i}\alpha_i^{\rm D};\;\forall i\in \Omega^D$, which can be replaced by the following continuous variables $z^G_{\varphi_i}$, $z^D_{\varphi_i}$, and $z^D_{\alpha_i}$, respectively, iff the following set of constraints is included:
%(\ref{lineaini})-(\ref{lineafin})
\begin{eqnarray}
  \varphi_i^{G,\rm min}z^G_{i} & \le & z^G_{\varphi_i} \le  \varphi_i^{G, \rm max} z^G_{i};\;\forall i\in \Omega^G \label{lineaini}\\
 \varphi_i^{G,\rm min}(1-z^G_{i}) & \le & \varphi_i^{\rm G}-z^G_{\varphi_i} \le  \varphi_i^{G, \rm max} (1-z^G_{i});\;\forall i\in \Omega^G \\
  \varphi_i^{D, \rm min}z^D_{i} & \le & z^D_{\varphi_i} \le  \varphi_i^{D, \rm max} z^D_{i};\;\forall i\in \Omega^D \\
 \varphi_i^{D, \rm min}(1-z^D_{i}) & \le & \varphi_i^{\rm D}-z^D_{\varphi_i} \le  \varphi_i^{D,\rm max} (1-z^D_{i});\;\forall i\in \Omega^D \\
  \alpha_i^{\rm min}z^D_{i} & \le & z^D_{\alpha_i} \le  \alpha_i^{\rm max} z^D_{i};\;\forall i\in \Omega^D \\
 \alpha_i^{\rm min}(1-z^D_{i}) & \le & \alpha_i^{\rm D}-z^D_{\alpha_i} \le  \alpha_i^{\rm max} (1-z^D_{i});\;\forall i\in \Omega^D,\label{lineafin}
\end{eqnarray}
where parameters $\varphi_i^{G,\rm min}$, $\varphi_i^{G,\rm max}$, $\varphi_i^{D,\rm min}$, $\varphi_i^{D,\rm max}$, $\alpha_i^{\rm min}$ and $\alpha_i^{\rm max}$ are the lower and upper limits of the dual variables, which can be replaced by $-M$ and $M$, respectively, with $M$ being a positive and large enough constant. Note that in comparison with the subproblem proposed by Jabr in \citep{Jabr:13}, we take advantage of the fact that we know in advance what uncertain variables tend, respectively, to the upper and lower limit of the uncertainty set. This strategy allows the number of binary variables to be halved.

Finally, the iterative scheme set out is described step by step on the following algorithm:
\begin{algorithm}{Iterative method}
\begin{description}
\item[Input:] Selection of uncertainty budgets $\Gamma_r^G$ and $\Gamma_r^D$ for each region and the tolerance of the  $\varepsilon$ process. This data is selected by the decision maker.

\item [Step 1: Initialization.] Initialize the iteration counter to $\nu=1$, and upper and lower limits of the objective function $z^{(\rm up)} = \infty$ and $z^{(\rm lo)} = -\infty$.

\item [Step 2: Solving the master problem at iteration $\nu$.] Solve the master problem (\ref{master5}) subject to constraints (\ref{master7})-(\ref{master12}). The result provides values of the decision variables ${\bfg x}_{(\nu)}$ and $\gamma_{(\nu)}$. Update the lower limit of the optimal objective function  $z^{(\rm lo)}={\bfg c}^T{\bfg x}_{(\nu)}+\gamma_{(\nu)}$. Note that at the first iteration the optimal solution corresponds to the no investment scenario, alternatively, we could start with any other vector for decision variables.

\item [Step 3: Solving subproblem at iteration $\nu$.] For given values of the decision variables ${\bfg x}_{(\nu)}$, we calculate the least optimum operating costs within the uncertainty set $f^{\rm dual}_{(\nu)}$, also obtaining  the corresponding uncertain parameters ${\bfg d}_{(\nu)}$. This is achieved by solving problem (\ref{subproblem1}) subject to constraints (\ref{dual2}), (\ref{dual3}), (\ref{dual4}) and the uncertainty set definition (\ref{uncer1})-(\ref{uncer2}).

    Update the upper limit of the optimal objective function upper bound $z^{(\rm up)}={\bfg c}^T{\bfg x}_{\nu}+f^{\rm dual}_{(\nu)}$.

\item [Step 4: Convergence checking.] If $(z^{(\rm up)}-z^{(\rm lo)})/z^{(\rm up)}\le \varepsilon$ go to {\em Step 5}, else update the iteration counter $\nu\rightarrow \nu+1$ and continue in {\em Step 2}.

\item [Step 5: Output.] The solution for a given tolerance corresponds to ${\bfg x}^\ast={\bfg x}_{(\nu)}$.

\end{description}
\end{algorithm}

\subsection{Difference in computational complexity}

To show the difference in computational complexity associated with the subproblems set out by Jabr \citep{Jabr:13}, by Cheng et al. \citep{ChenWWHW:14} and Ruiz and Conejo \citep{RuizC:14}, and that set out in this paper (\ref{subproblem1})-(\ref{subproblem2}), we calculate the number of variables and constraints required for each formulation in terms of the number of generators ($n_g$), demand ($n_d$), buses ($n_b$) and branches ($n_l$):
\begin{description}
  \item[Jabr subproblem:] In this case the number of variables and constraints is equal to:
  \begin{enumerate}
  \item $2(n_g+n_d)$ binary variables associated with uncertain parameters. Note that each pair of binary variables corresponds to values above and below the nominal value, respectively.
  \item $4n_g+6n_d+3n_b+3n_l$ continuous variables associated with the dual of the third-level problem, which optimizes operational cost.
  \item $6n_g+11n_d+n_l+n_b+3$ constraints associated with the third-level dual problem, objective function, robust budgets related to demand and maximum power generation, and the linearization formulas.
\end{enumerate}
  \item[Cheng et al. and Ruiz and Conejo subproblem:] In this case the number of variables and constraints is equal to:
  \begin{enumerate}
%%%  \item $n_g+n_d$ binary variables associated with uncertain parameters.
  \item $2n_g+2n_d+2(n_b-1)+2n_l$ binary variables related to the Fortuny-Amat transformation.
\item $5n_g+7n_d+4n_b+4n_l$ continuous variables associated with the primal and dual of the third-level problem, which optimizes operational costs.
  \item $3n_g+5n_d+4n_l+4n_b$ constraints related to the third-level primal and dual problem, objective function, and robust budgets related to demand and maximum power generation.
  \item $4n_g+4n_d+4n_l+4n_b-2$ constraints associated with the Fortuny-Amat transformation.
\end{enumerate}
\item[Proposed subproblem:] In this case the number of variables and constraints is equal to:
  \begin{enumerate}
  \item $n_g+n_d$ binary variables associated with uncertain parameters.
  \item $3n_g+5n_d+3n_b+3n_l$ continuous variables associated with the dual of the third-level problem, which optimizes operational cost.
  \item $5n_g+10n_d+n_l+n_b+3$ constraints associated with the third-level dual problem, objective function, robust budgets related to demand and maximum power generation, and the linearization formulas.
      \end{enumerate}
\end{description}

According to these figures, and considering that the higher number of elements in real networks are lines and buses, it is clear that the approach using the KKT conditions increases the number of binary variables with respect to other methods by almost doubling the number of lines and buses. Moreover, the number of constraints is also considerably higher. This is why the computational cost for subproblems increases substantially. If we compare the formulation proposed by Jabar and the proposal in this paper, the number of variables reduces by half for the latter, thereby reducing the computational time required to solve the proposed formulation.

Note that we do not compare computational complexity for master problems because it has been demonstrated in the current literature that the column and constraint generation method is more efficient than Benders decomposition.

\section{Numerical case studies}\label{CaseStudy}
In this section, we set out numerical experiments of our model and compare them with the method proposed in \citep{ChenWWHW:14} and \cite{RuizC:14}. We use an illustrative example i.e. the Garver system \citep{Garver:70}, and two realistic case studies: the IEEE 24-bus system \citep{IEEE_RTS:99} and the IEEE 118-bus test system \cite{PSTCA:13}. It must be stressed that in this paper we just focus on numerical performance as interpretation of robust solutions is beyond its scope  but in references \cite{Jabr:13}, \citep{ChenWWHW:14} and \cite{RuizC:14}further details are given.

All examples have been implemented and solved using GAMS \citep{GAMS,BrookeKMR:98} and CPLEX 12, on a PC with four processors clocking at 2.39 GHz and 3.2 GB of RAM memory. Note that computing times reported correspond to the total running time used by the solvers to work out the masters and subproblems until the final solution is attained. We use the same tolerance for all problems that are equal to $\varepsilon=10^{-6}$.

\subsection{Illustrative example. Garver system}
\begin{figure}[htb]
  \begin{center}
  \includegraphics[width=0.5\textwidth]{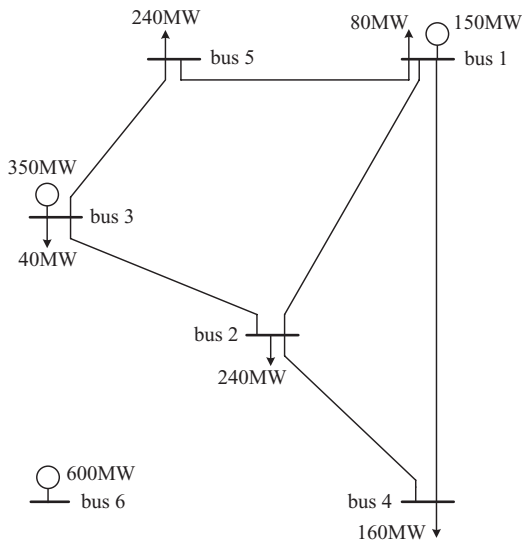}
    \caption{Garver's 6-bus test system.}
    \label{Figure1}
  \end{center}
\end{figure}

The model set out is illustrated with the Garver 6-bus system, depicted in Figure~\ref{Figure1}. This system is made up of 6 buses, 3 generators, 5 types of inelastic demand and 6 lines. Nominal values of generation capacities and demand and their prices and nominal costs can be found in Table~\ref{data garver}. The load-shedding cost is equal to the nominal cost of each demand type.
\begin{table}[htb]
\renewcommand{\arraystretch}{1.2}
\renewcommand{\tabcolsep}{2.0pt}
{\footnotesize
\begin{center}
\caption{Generator and demand nominal values for Garver's 6-bus example}\label{data garver}
\begin{tabular}{|c|c|c|c|c|}\hline
& \multicolumn{2}{|c|}{Generators}& \multicolumn{2}{|c|}{Demands}\\\hline
Bus & \begin{tabular}{c} Power size \\ (MW)\end{tabular} &  \begin{tabular}{c} Cost \\ (\euro/MWh)\end{tabular}  &
\begin{tabular}{c} Load \\ (MW)\end{tabular} &  \begin{tabular}{c} NLS cost \\ (\euro/MWh)\end{tabular} \\ \hline
1 & 150 & 60 & 80 & 112.5\\ \hline
2 &  -- & -- & 240 & 115\\ \hline
3 & 350 & 65 & 40 & 120 \\ \hline
4 & -- & -- & 160 & 110\\ \hline
5 & -- &  --&240 & 112\\ \hline
6 & 600 &  70 & -- &--\\ \hline
\multicolumn{5}{l}{NLS: Nominal load-shedding}\\
 \hline
\end{tabular}
\end{center}
}
\end{table}
Let us consider that a maximum of three lines can be installed between each pair of buses. Line data is obtained from Table I of \cite{GarcesCGR:09} including construction costs, and the maximum funds available for investment is 40 million euros.

The return on investment period for each line is considered to be 25 years, and the discount rate is 10\%, which results in an annual amortization rate of 11\%. Since the investment cost is annualized, the weighted factor $\sigma$ is equal to the number of hours in a year, i.e. 8760, by which an annualized load-shedding and power generation cost can be obtained.

Regarding the uncertainty set, power generation capacity can oscillate by a maximum of 50\% of their nominal values, while demand can change up to a maximum of 20\%.

We have solved the robust TNEP problem for three different combinations of uncertainty budgets associated with generation capacities and demand. We have used both the method proposed by \citep{ChenWWHW:14} and \cite{RuizC:14} and that set out in this paper. In order to obtain statistically sound conclusions about computing times, we have solved the problem 100 times and obtained mean and standard deviation executions times. The solution times from these numerical experiments are given in Table~\ref{single_garver}.

\begin{table}[htb]
\renewcommand{\arraystretch}{1.2}
\renewcommand{\tabcolsep}{2.0pt}
{\footnotesize
\begin{center}
\caption{Computational results for Garver's 6-bus example}\label{single_garver}
\begin{tabular}{|c|c|c|c|c|c|c|}
\hline
\multicolumn{3}{|c|}{} & \multicolumn{2}{|c|}{\citep{ChenWWHW:14} and \cite{RuizC:14} method} & \multicolumn{2}{|c|}{Proposed method}\\\hline
 &  \begin{tabular}{c} Optimal \\  sol.  (M\euro)\end{tabular} & $\sharp$ iter. & Mean (s) & Std. (s) & Mean (s) & Std. (s) \\
\hline
 \begin{tabular}{c} $\Gamma^{\rm G}=3$ \\ $\Gamma^{\rm D}=5$\end{tabular} &  35505.31 & 3 & 0.973 & 0.048 & 0.644 & 0.054\\ \hline
 \begin{tabular}{c} $\Gamma^{\rm G}=2$ \\ $\Gamma^{\rm D}=3$\end{tabular} &  25832.22 & 4 & 1.339 & 0.106 & 0.936 & 0.094\\ \hline
 \begin{tabular}{c} $\Gamma^{\rm G}=1$ \\ $\Gamma^{\rm D}=2$\end{tabular}  & 5861.92 & 4 & 1.605 & 0.095 & 1.011 & 0.072\\ \hline
 \begin{tabular}{c} $\Gamma^{\rm G}=0$ \\ $\Gamma^{\rm D}=0$\end{tabular}  & 440.07 & 2 & 0.539 & 0.071 & 0.428 & 0.065\\ \hline
\end{tabular}
\end{center}
}
\end{table}

From Table~\ref{single_garver} the following observations are pertinent:
\begin{enumerate}
  \item The method set out is computationally faster on average than that put forward by \citep{ChenWWHW:14} and \cite{RuizC:14} for all situations, from the least desirable case which corresponds to Soyster's solution \citep{Soyster:73} through to the deterministic case when uncertainty budgets are null. Regarding the standard deviation of computing times the values are comparable. Computational time in the method set out in this paper is approximately 30\% less than the other method with which it has been compared.

  \item Both methods provide the same solution and converge in the same number of iterations. Note that if the uncertainty budgets are integer, both approaches are equivalent because the master problem is the same and the subproblems are equivalent.

  \item The maximum number of iterations required is four.
\end{enumerate}

The difference in computing time is associated with the solution to the subproblem. That set out by \citep{ChenWWHW:14} and \cite{RuizC:14} has 472 equations, 245 continuous and 119 binary variables, while the subproblem set out in this paper contains 119 equations, 187 continuous and 8 binary variables. Even though we are dealing with a small example, the difference in complexity between both subproblems is considerable, especially in terms of binary variables.

\subsection{IEEE 24-bus Reliability Test System}
\begin{figure}[htb]
  \begin{center}
  \includegraphics[width=0.6\textwidth]{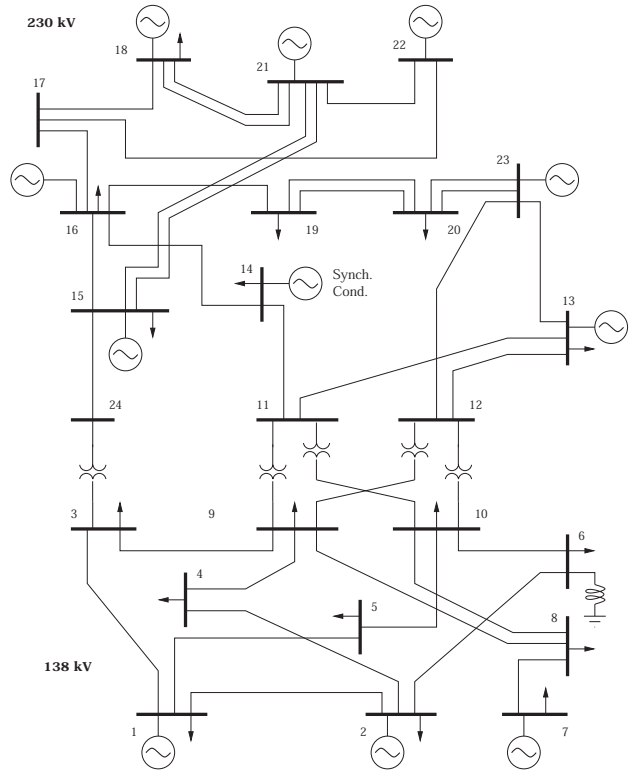}
    \caption{IEEE 24-bus reliability test system (RTS)}
    \label{Figure2}
  \end{center}
\end{figure}

The following case study is based on the IEEE 24-bus Reliability Test System (RTS) \citep{IEEE_RTS:99}, depicted in Fig.~\ref{Figure2}. The system comprises 24 buses, 34 existing corridors which can accept a maximum of three equal lines and 7 new corridors, $10$ generating units and $17$ loads. Data for lines in existing corridors has been taken from \citep{IEEE_RTS:99}, while line data for new corridors has been obtained from Table I in \citep{FangH:03}. Investment costs are 20 million euros, and analogously to the Garver illustrative example, the  return on investment period for each line is considered to be 25 years, and the discount rate is 10\%, which results in an annual amortization rate of 10\%.
Table \ref{busdata24} shows where generators and demand can be located in the network, the maximum power and the corresponding costs for generating units, and the load and the nominal load-shedding cost for demand levels. The load-shedding cost is equal to 100 times the nominal value of each level of demand.
\begin{table}[htb]
\renewcommand{\arraystretch}{1.0}
\renewcommand{\tabcolsep}{1.0pt}
{\footnotesize
\begin{center}
\caption{Generator and demand data for IEEE 24 RTS case study}\label{busdata24}
\begin{tabular}{|c|c|c|c|c|}\hline
& \multicolumn{2}{c|}{Generators} & \multicolumn{2}{c|}{Demands} \\ \hline
\multirow{2}{*}{Bus} & Power size & Cost & Load & NLS cost \\
& (MW) & (\euro/MWh) & (MW) & (\euro/MWh) \\ \hline

Bus & \begin{tabular}{c} Power size \\ (MW)\end{tabular} &  \begin{tabular}{c} Cost \\ (\euro/MWh)\end{tabular}  &
\begin{tabular}{c} Load \\ (MW)\end{tabular} &  \begin{tabular}{c} NLS cost \\ (\euro/MWh)\end{tabular} \\ \hline
1 & 230 & 95 & 259 & 99 \\ \hline
2 & 230 & 96 & 233 & 98 \\ \hline
3 & --&--& 432 & 100 \\ \hline
4 &--&--& 178 & 99 \\ \hline
5 &--&--& 171 & 100 \\ \hline
6 &--&--& 326 & 99 \\ \hline
7 & 360 & 96 & 300 & 100 \\ \hline
8 &--&--& 411 & 93\\ \hline
9 &--&--& 420 & 99 \\ \hline
10 &--&--& 468 & 100\\ \hline
%11 &--&--&--& \\ \hline
%12 &--&--&--& \\ \hline
13 & 709 & 80 & 636 & 92\\ \hline
14 &--&--& 466 & 90\\ \hline
15 & 258 & 82 & 761 & 87\\ \hline
16 & 186 & 77 & 240 & 84\\ \hline
%17 &--&--&--& \\ \hline
18 & 480 & 73 & 799 & 91\\ \hline
19 &--&--& 435 & 94\\ \hline
20 &--&--& 307 & 95 \\ \hline
21 & 480 & 74 &--&-- \\ \hline
22 & 360 & 79 &--& --\\ \hline
23 & 792 & 78 &--&-- \\ \hline
24 &--&--&--&-- \\ \hline
\multicolumn{5}{l}{NLS: Nominal load-shedding}\\
 \hline
\end{tabular}
\end{center}
}
\end{table}

Regarding the uncertainty set, power generation capacity can oscillate by a maximum of 50\% of its nominal values, while demand can change by up to a maximum of 20\%.

We have solved the same computational experiment as in the previous illustrative example, also comparing it with the method set out by \citep{ChenWWHW:14} and \cite{RuizC:14}. Solution times from these numerical experiments are given in Table~\ref{single_24RTS}.

\begin{table}[htb]
\renewcommand{\arraystretch}{1.2}
\renewcommand{\tabcolsep}{2.0pt}
{\footnotesize
\begin{center}
\caption{Computational results for IEEE 24 RTS case study}\label{single_24RTS}
\begin{tabular}{|c|c|c|c|c|c|c|}
\hline
\multicolumn{3}{|c|}{} & \multicolumn{2}{|c|}{\citep{ChenWWHW:14} and \cite{RuizC:14} method} & \multicolumn{2}{|c|}{Proposed method}\\\hline
 &  \begin{tabular}{c} Optimal \\  sol.  (M\euro)\end{tabular} & $\sharp$ iter. & Mean (s) & Std. (s) & Mean (s) & Std. (s) \\
\hline
 \begin{tabular}{c} $\Gamma^{\rm G}=10$ \\ $\Gamma^{\rm D}=17$\end{tabular} &  500752.14 & 3 & 165.64 & 46.63 & 1.51 & 0.30 \\ \hline
 \begin{tabular}{c} $\Gamma^{\rm G}=7$ \\ $\Gamma^{\rm D}=12$\end{tabular} &  454917.47 & 3 & 308.20 & 75.97 & 1.80 & 0.39\\ \hline
 \begin{tabular}{c} $\Gamma^{\rm G}=3$ \\ $\Gamma^{\rm D}=5$\end{tabular}  & 348941.40 & 3 & 518.75 & 1.15 & 1.12 & 0.11\\ \hline
 \begin{tabular}{c} $\Gamma^{\rm G}=0$ \\ $\Gamma^{\rm D}=0$\end{tabular}  & 219161.52 & 2 & 144.29 & 0.17 & 0.414 & 0.08\\ \hline
\end{tabular}
\end{center}
}
\end{table}

From Table~\ref{single_24RTS} we can extract the same conclusions as in the previous illustrative example:
\begin{enumerate}
\item The  method set out is considerably faster in terms of average computing time than that described by \citep{ChenWWHW:14} and \cite{RuizC:14} for every situation, from the least desirable scenario which corresponds to Soyster's solution \citep{Soyster:73} through  to the deterministic case, with the latter being that having the lowest computing times. This is because the values of the binary variables of the subproblems are known in advance and only 2 iterations are required. This has made the average computational time 235 times faster.
  \item The maximum number of iterations required is three.
\end{enumerate}

The difference in computing time is associated with the solution to the subproblem. That set out by \citep{ChenWWHW:14} and \cite{RuizC:14} has 1397 equations, 757 continuous and 346 binary variables, while the subproblem set out in this paper contains 370 equations, 556 continuous and 27 binary variables. The difference in complexity between both subproblems is considerable.

\subsection{IEEE 118-bus test system}
FFinally, we ran additional computational tests using the IEEE 118-bus test system~\citep{PSTCA:13}. The system was made up of 118 buses, 186 existing lines, $54$ generating units and $91$ loads. In addition, it was possible to construct up to 61 additional lines to duplicate each one of the following existing lines: 8, 12, 23, 32, 38, 41, 51, 68, 78, 96, 104, 118, 119, 121, 125, 129, 134, 159, 7, 9, 36, 117, 71, 131, 133, 147, 103, 65, 144, 168, 4, 13, 132, 69, 66, 67, 5, 89, 29, 167, 145, 70, 42, 90, 16, 174, 98, 99, 185, 93, 94, 128, 164, 97, 153, 146, 116, 163, 31, 92, 130. Data for lines in existing corridors was taken from \citep{PSTCA:13}. The investment cost came to  100 million euros, and as in both previous examples, the return on investment  period for each line was considered to be 25 years, and the discount rate was 10\%, which resulted in a 10\% annual amortization rate.

Table~\ref{busdata118} in~\ref{IEEE-118data} shows where the generators and demand is located in the network, the maximum power and the corresponding cost for generating units, and the load and nominal load-shedding cost for demand levels. The load-shedding cost is equal to 1.2 times the nominal load-shedding cost for each level of demand.

Regarding the uncertainty set, power generation capacity and demand can oscillate up to a maximum of 50\% of their nominal values.

We have solved the same computational experiment as in the previous illustrative example, and have compared it with the method set out by \citep{ChenWWHW:14} and \cite{RuizC:14}. The solution times from these numerical experiments are given in Table~\ref{single_118}.

\begin{table}[htb]
\renewcommand{\arraystretch}{1.2}
\renewcommand{\tabcolsep}{2.0pt}
{\footnotesize
\begin{center}
\caption{Computational results for IEEE 118-bus test system}\label{single_118}
\begin{tabular}{|c|c|c|c|c|c|c|}
\hline
\multicolumn{3}{|c|}{} & \multicolumn{2}{|c|}{\citep{ChenWWHW:14} and \cite{RuizC:14} method} & \multicolumn{2}{|c|}{Proposed method}\\\hline
 &  \begin{tabular}{c} Optimal \\  sol.  (M\euro)\end{tabular} & $\sharp$ iter. & Mean (s) & Std. (s) & Mean (s) & Std. (s) \\
\hline
 \begin{tabular}{c} $\Gamma^{\rm G}=54$ \\ $\Gamma^{\rm D}=91$\end{tabular} &  31994.36 & 4 & 36000$^\ast$ & -- & 7.734 & 0.800 \\ \hline
 \begin{tabular}{c} $\Gamma^{\rm G}=35$ \\ $\Gamma^{\rm D}=60$\end{tabular} &  30032.93 & 5 & 36000$^\ast$ & -- & 53.460 & 60.360\\ \hline
 \begin{tabular}{c} $\Gamma^{\rm G}=15$ \\ $\Gamma^{\rm D}=20$\end{tabular}  & 23352.97 & 4 & 36000$^\ast$ & -- & 24.260 & 26.050\\ \hline
 \begin{tabular}{c} $\Gamma^{\rm G}=0$ \\ $\Gamma^{\rm D}=0$\end{tabular}  & 13929.23 & 2 & 36000$^\ast$ & -- & 0.560 & 0.073\\ \hline
\end{tabular}
\end{center}
}
\end{table}

From Table~\ref{single_118} the following observations are pertinent:
\begin{enumerate}
  \item The method set out in this paper is far faster in average computing time than that proposed in \citep{ChenWWHW:14} and \cite{RuizC:14} for every situation. In fact, with the latter method the optimal solution was not reached within the set time limit of 36000 s (10 hours). Note that if any of the solvers (for master or subproblems) does not find an optimal solution within that time the process is stopped. Conversely, with the method set out in this paper the optimal solution was reached in less than one minute for all instances of budget uncertainties.

  \item The maximum number of iterations required was 5.

\end{enumerate}

The difference in computing time is associated with the solution to the subproblem. That set out by \citep{ChenWWHW:14} and \cite{RuizC:14} has 4115 equations, 2367 continuous and 1018 binary variables, while the subproblem set out in this paper contains 1548 equations, 1712 continuous and 145 binary variables. The difference in complexity between both subproblems is apparent and explains the large difference in computing time between both methods.

\subsection{Discussion}
Numerical experiments demonstrate that the method set out in this paper is computationally more efficient than that by \citep{ChenWWHW:14} and \cite{RuizC:14}. Nevertheless, this result is evident considering the difference in complexity of the subproblem. Note that Ruiz and Conejo \cite{RuizC:14} already claimed that the most time-consuming stage of their process was the subproblem resolution, and it is the subproblems solved by our method that really makes a difference in terms of saving computational time, while master problems remain unchanged.

We have not reported results with respect to the method proposed by Jabr \citep{Jabr:13}. Nevertheless, it is also evident that our method is more efficient because we use the same subproblem whilst halving the number of binary variables. This difference might imply large computing time differences for large problems. In fact, the larger the problem, the larger the difference which is due to how standard branch-and-cut solvers function. The only uncertainty as to whether our method is faster might stem from the master problem. However, the number of iterations reported in this paper and in references \citep{ChenWWHW:14} and \cite{RuizC:14} are always lower than or equal to 5, while the number of iterations reported by \citep{Jabr:13} varied considerably depending if additional cuts were included or not, but they were usually above 5 or more. Furthermore, this strategy of including additional primal cuts could be used in our master problem by just considering additional outcomes for uncertain parameters as Jabr \citep{Jabr:13} does in his work. Nevertheless, because of the convergence behavior this is not necessary.

\section{Conclusions}\label{Conclusions}

In this paper a new decomposition algorithm has been set out to attain the exact solution for the TNEP problem stemming from the use of a two-stage adaptive robust strategy. Although there is nothing novel with respect to the formulation of the problem, we have managed to combine formulations and findings from different authors in the proper manner. Therefore, in terms of computational efficiency, our method leads, occasionally by a large margin, than other methodologies in use.

We have run several computational experiments to show how the  algorithm set out in this paper functions with respect to the approach set out by \citep{ChenWWHW:14} and \cite{RuizC:14}. Nevertheless, it has been demonstrated that the computational time required by this algorithm is always lower than those set out by \citep{Jabr:13}, \citep{ChenWWHW:14} and \citep{RuizC:14}. The subproblem set out in this paper is less complex and has fewer binary variables than as is the case in other research, while the primal master problem is equal to that set out by \citep{ChenWWHW:14} and \citep{RuizC:14}, which is at an advantage in terms of computing with respect to the Benders master problem used in \citep{Jabr:13}, and, moreover, the inclusion of additional cuts from given realizations of the random parameters is not required.

Bearing in mind that with the method set out all features as regards robust optimization design with respect to their predecessors were maintained, the algorithm set out in this paper turned out to be the most efficient method for solving the robust transmission network expansion planning to date, and takes advantage of state-of-the-art mixed-integer mathematical programming solvers. The only limitation is the integrality constraint of budget uncertainty, which simplifies the subproblem considerably although without detracting from the benefits of using robust optimization.

\section*{Acknowledgements}
This work set out in this paper has been partially funded by projects by the Junta de Comunidades of Castilla-La Mancha, under project POII-2014-012-P; the Ministry of Science of Spain, under CICYT Project ENE2012-30679; and by the European Commission, under Grant Agreement Number 309048.

\appendix

\section{Detailed formulation of the TNEP problem}\label{detfor}

Before describing the detailed formulation, the main notation used and not defined previously is stated below for quick
reference.\\

{\bf Constants:}
\begin{description}

\item[$b_{k}$] Susceptance of line $k$ (S).

\item[$c^{\rm G}_{i}$] Generation cost for generator $i$ (\euro/MWh).

\item[$c^{\rm U}_{j}$] Load-shedding cost for consumer $j$ (\euro/MWh).

\item[$c_{k}$] Investment cost of building line $k$ (\euro).

\item[$e_j$] Percentage of load shed by the $j$-th demand.

\item[$f_{k}^{\rm max}$] Capacity of line $k$ (MW).

\item[$o(k)$] Sending-end bus of line $k$.

\item[$r(k)$] Receiving-end bus of line $k$.

\end{description}

{\bf Primal variables:}
\begin{description}
\item[$d_{j}$] Power consumed by the $j$-th demand (MW).

\item[$f_{k}$] Power flow through line $k$ (MW).

\item[$g_{i}$] Power produced by the $i$-th generating unit  (MW).

\item[$r_{j}$] Load shed by the $j$-th demand (MW).

\item[$x_{k}$] Binary variable that is equal to $1$ if line $k$ is built and $0$ otherwise.

\item[$\theta_{s}$] Voltage angle at bus $s$ (radians).

\end{description}

{\bf Dual variables:} Note that dual variables are provided after the corresponding equalities or inequalities separated by a colon.\\

{\bf Indices and Sets:}
\begin{description}
%
%[\setlabelwidth{$g_{ib}^{\rm max}(w)$}\usemathlabelsep]
%
%
\item[$s(i)$] Bus index where the $i$-th generating unit is located.

\item[$s(j)$] Bus index where the $j$-th demand is located.

\item[$\Psi_{s}^{\rm D}$] Set of indices of the demand located at bus $s$.

\item[$\Psi_{s}^{\rm G}$] Set of indices of the generating units located at bus $s$.

\item[$\Omega^{\rm D}$] Set of indices of the demand.

\item[$\Omega^{\rm G}$] Set of indices of the generating units.

\item[$\Omega^{\rm L}$] Set of all prospective and existing transmission lines.

\item[$\Omega^{\rm L^{\rm +}}$] Set of all prospective transmission lines.

\item[$\Omega^{\rm N}$] Set of all networks buses.

\end{description}

{\bf TNEP detailed problem:}\\

The decision-making problem pertaining to a transmission planner simultaneously minimizes network investment, generation and load-shedding costs. Problem (\ref{eq1})-(\ref{eq3}) for given values of the random parameters ${\bfg d}$ is as follows:
\begin{equation}\label{OF1}
\mm{Minimize}{{\bfg x},{\bfg y}}{\displaystyle R\left(\sum_{k\in \Omega^{\rm L^{\rm +}}}c_{k}x_{k}\right)+\sigma\left(\sum_{i\in \Omega^{\rm G}}c^{\rm G}_{i}g_{i}+\sum_{j\in \Omega^{\rm D}}c^{\rm U}_{j}r_{j}\right)}
\end{equation}
subject to
\begin{eqnarray}
x_{k}&=&1; \hspace{2mm}  \forall k \in \Omega^{\rm L} \backslash \Omega^{\rm L^{\rm +}} \label{exist lin} \\
x_{k}& \in& \{0,1\}; \hspace{2mm}  \forall k\in \Omega^{\rm L} \label{decision var} \\
\displaystyle\sum_{k\in \Omega^{\rm L^{\rm +}}}c_{k}x_{k}& \leq& \Pi \label{invest max} \\
\sum_{i \in \Psi_{s}^{\rm G}}g_{i}-\sum_{k \mid o(k)=s}f_{k} &&\nonumber\\
+\sum_{k \mid r(k)=s}f_{k}+\sum_{j \in \Psi_{s}^{\rm D}}r_{j} & = & \sum_{j\in \Psi_{s}^{\rm D}}d_{j}: \lambda_{s}; \hspace{2mm} \forall s \in \Omega^{\rm N}
\label{balance}\\
f_{k}&=&b_{k}x_{k}(\theta_{o(k)}-\theta_{r(k)}): \phi_{k};\hspace{2mm} \forall k \in \Omega^{\rm L} \label{flow}\\
\theta_{s}&=&0: \chi_{s}; \hspace{2mm} s:\text{slack} \label{refer angle}\\
\end{eqnarray}
\begin{eqnarray}
f_{k} &\leq& f_{k}^{\rm max}: \phi_{k}^{\rm max}; \hspace{2mm} \forall k \in \Omega^{\rm L}\label{flow upper limit}\\
f_{k}&\geq& -f_{k}^{\rm max}: \phi_{k}^{\rm min}; \hspace{2mm} \forall k \in \Omega^{\rm L}\label{flow lower limit}\\
\theta_{s} & \leq & \pi: \xi_{s}^{\rm max}; \quad
\forall s \in \Omega^{\rm N}\backslash s:\text{slack} \label{angle upper limit}\\
\theta_{s}& \geq &-\pi: \xi_{s}^{\rm min}; \quad
\forall s \in \Omega^{\rm N}\backslash s:\text{slack} \label{angle lower limit}\\
%
%
%d_{j} &\geq& 0; \hspace{2mm}
%\forall j \in \Omega^{\rm D} \label{demand lower limit}\\
%
%
g_{i} &\geq& 0; \hspace{2mm}
\forall i\in\Omega^{\rm G}  \label{gener lower limit}\\
r_{j}& \geq &0; \hspace{2mm}
\forall j \in \Omega^{\rm D}\label{loadshed lower limit}\\
d_{j} &=& d_{j}^{\rm D}: \alpha_{j}^{\rm D}; \hspace{2mm}
\forall j \in \Omega^{\rm D}\label{demand upper limit}\\
g_{i}& \leq& d_{i}^{\rm G}: \varphi_{i}^{\rm G}; \hspace{2mm}
\forall i\in\Omega^{\rm G}  \label{gener upper limit}\\
r_{j} &\leq& e_j d_{j}^{\rm D}: \varphi_{j}^{\rm D}; \hspace{2mm} \forall j\in \Omega^{\rm D}. \label{loadshed upper limit}
\end{eqnarray}

The generation and load-shedding costs are multiplied by the weighting factor $\sigma$ to make the annual investment cost and the generation and load-shedding cost comparable quantities. Investment costs in (\ref{OF1}) are multiplied by the capital recovery factor $R$, calculated as $\fraca{r(1+r)^n}{(1+r)^n-1}$, where $r$ is the interest rate and $n$ is the number of  years considered. Note that dual variables are provided after the corresponding equalities or inequalities separated by a colon. Constraint (\ref{invest max}) sets an upper limit on the investment cost and corresponds to (\ref{eq2}). Constraints (\ref{balance}) set the power balance at every bus. Constraints (\ref{flow}) represent the power flow through each line. Each of these constraints is multiplied by a binary variable, thus, if the corresponding line is not physically connected to the network, the power flow through it is zero. Equation (\ref{refer angle}) fixes the voltage angle of the reference bus. Note that (\ref{balance})-(\ref{refer angle}) corresponds to the linear constraint set associated with $\Omega({\bfg x},{\bfg d})$ in the first row of (\ref{eq4}).

Constraints (\ref{flow upper limit})-(\ref{flow lower limit}) set the line flow limits. Constraints (\ref{angle upper limit})-(\ref{angle lower limit}) set limits on the voltage angles at every bus, and (\ref{gener lower limit})-(\ref{loadshed lower limit}) ensures  power generation and load-shedding are both positive. All these inequality constraints corresponds to the inequality set related to $\Omega({\bfg x},{\bfg d})$ in the second row of (\ref{eq4}).

Constraints (\ref{demand upper limit}) force demand levels to be equal to the uncertain demand variable, and corresponds to the equality constraint set related to $\Omega({\bfg x},{\bfg d})$ in the third row of (\ref{eq4}). Finally, (\ref{gener upper limit}) and (\ref{loadshed upper limit}) set power generation and load-shedding to be lower than the generation capacity and a percentage of  demand, respectively. These last two sets of constraints correspond to the inequality constraint set related to $\Omega({\bfg x},{\bfg d})$ in the fourth row of (\ref{eq4}).\\

{\bf Detailed subproblem:}\\

For the sake of completeness, we also provide the detailed formulation of the proposed subproblem, made up of the objective function (\ref{subproblem1}) subject to constraints (\ref{dual2}), (\ref{dual3}), (\ref{dual4}), the uncertainty set definition (\ref{uncer1})-(\ref{uncer2}) and the linearization (\ref{lineaini})-(\ref{lineafin}). It is as follows:
\begin{equation}\label{OFdual}
\mm{Maximize}{{\bfg d},{\bfg \lambda},{\bfg \mu},{\bfg \alpha},{\bfg \varphi}}{\left\{
\begin{array}{c}
 \sum_{k \in \Omega^{\rm L}}\Bigl(\phi_{k}^{\rm max}-\phi_{k}^{\rm min}\Bigr)f_{k}^{\rm max}  +\sum_{s\in \Omega^{\rm N}\backslash s:\text{slack}}\pi\Bigl(\xi_{s}^{\rm max}-\xi_{s}^{\rm min}\Bigr) \\
 +\sum_{i\in \Omega^{\rm G}}\Bigl(\bar d^G_{i}\varphi_i^{\rm G}-\hat d^G_{i} z^G_{\varphi_i}\Bigr) +\sum_{j\in \Omega^{\rm D}}\Bigl(\bar d^D_{i}\alpha_i^{\rm D}+\hat d^D_{i} z^D_{\alpha_i}\Bigr)\\
  +\sum_{j\in \Omega^{\rm D}}e_j\Bigl(\bar d^D_{j}\varphi_j^{\rm D}+\hat d^D_{j}z^D_{\varphi_j}\Bigr)
 \end{array}\right\}
}
\end{equation}
subject to:
\begin{eqnarray}
\lambda_{s(i)}+\varphi_{i}^{\rm G}& \leq & c^{\rm G}_{i}\sigma; \hspace{2mm}
\forall i\in \Omega^{\rm G} \label{dual blockgen}\\
-\lambda_{s(j)}+\alpha_{j}^{\rm D} & \leq & 0; \hspace{2mm}
\forall j\in \Omega^{\rm D} \label{dual blockdemand}\\
\lambda_{s(j)}+\varphi_{i}^{\rm D} & \leq & c^{U}_{j}\sigma; \hspace{2mm}
\forall j\in \Omega^{\rm D} \label{dual loadshed}\\
-\lambda_{o(k)}+\lambda_{r(k)}+\phi_{k}+\phi_{k}^{\rm max}
+\phi_{k}^{\rm min} & = & 0; \hspace{2mm} \forall k\in \Omega^{\rm L} \label{dual flow}\\
-\sum_{k\mid o(k)=s}b_{k}x_{k}\phi_{k}+\sum_{k\mid r(k)=s}b_{k}x_{k}\phi_{k} &  & \nonumber \\
+\xi_{s}^{\rm max}+\xi_{s}^{\rm min} & = & 0 \hspace{2mm}
\forall s \in \Omega^{\rm N}\backslash s:\text{slack} \label{dual angle}\\
-\sum_{k\mid o(k)=s}b_{k}x_{k}\phi_{k}+\sum_{k\mid r(k)=s}b_{k}x_{k}\phi_{k}
 &  & \nonumber
 \end{eqnarray}
\begin{eqnarray}
+\chi_{s} & = & 0 \hspace{2mm}
s:\text{slack} \label{dual anglerefer}\\
-\infty \leq & \lambda_{s} & \leq\infty; \hspace{2mm}\forall s \in \Omega^{\rm N} \label{dual lambda}\\
-\infty \leq& \phi_{k} & \leq\infty; \hspace{2mm}\forall k \in \Omega^{\rm L} \label{dual phik}\\
-\infty \leq& \chi_{s}& \leq\infty; \hspace{2mm}
s:\text{slack} \label{dual chi}
 \end{eqnarray}
\begin{eqnarray}
 \phi_{k}^{\rm max}&\leq& 0; \hspace{2mm}
\forall k  \in \Omega^{\rm L}\label{dual phikmax}\\
 \phi_{k}^{\rm min}&\geq& 0; \hspace{2mm}
\forall k \in \Omega^{\rm L}\label{dual phikmin}
\end{eqnarray}
\begin{eqnarray}
 \xi_{s}^{\rm max}&\leq& 0; \hspace{2mm}
\forall s  \in \Omega^{\rm N}\backslash s:\text{slack}\label{dual ximax}\\
 \xi_{s}^{\rm min}&\geq& 0; \hspace{2mm}
\forall s \in \Omega^{\rm N}\backslash s:\text{slack}\label{dual ximin}
 \end{eqnarray}
\begin{eqnarray}
-\infty \leq & \alpha_{j}^{\rm D} & \leq\infty; \hspace{2mm}\forall j \in \Omega^{\rm D} \label{dual alpha}\\
\varphi_{i}^{\rm G}& \leq &0; \hspace{2mm}
\forall i  \in \Omega^{\rm G}\label{dual varphiG}\\
\varphi_{i}^{\rm D}& \leq &0; \hspace{2mm}
\forall i  \in \Omega^{\rm D},\label{dual varphiD}
\end{eqnarray}
\begin{eqnarray}
  \varphi_i^{G,\rm min}z^G_{i} & \le & z^G_{\varphi_i} \le  \varphi_i^{G, \rm max} z^G_{i};\;\forall i\in \Omega^G \label{lineaini2}\\
 \varphi_i^{G,\rm min}(1-z^G_{i}) & \le & \varphi_i^{\rm G}-z^G_{\varphi_i} \le  \varphi_i^{G, \rm max} (1-z^G_{i});\;\forall i\in \Omega^G \\
  \varphi_i^{D, \rm min}z^D_{i} & \le & z^D_{\varphi_i} \le  \varphi_i^{D, \rm max} z^D_{i};\;\forall i\in \Omega^D \\
 \varphi_i^{D, \rm min}(1-z^D_{i}) & \le & \varphi_i^{\rm D}-z^D_{\varphi_i} \le  \varphi_i^{D,\rm max} (1-z^D_{i});\;\forall i\in \Omega^D \\
  \alpha_i^{\rm min}z^D_{i} & \le & z^D_{\alpha_i} \le  \alpha_i^{\rm max} z^D_{i};\;\forall i\in \Omega^D \\
 \alpha_i^{\rm min}(1-z^D_{i}) & \le & \alpha_i^{\rm D}-z^D_{\alpha_i} \le  \alpha_i^{\rm max} (1-z^D_{i});\;\forall i\in \Omega^D.\label{lineafin2}
\end{eqnarray}

Note that this is a mixed-integer linear mathematical programming formulation. Finally, the correspondence of dual variables between the compact and detailed formulation is as follows:
\begin{eqnarray}
  {\bfg \lambda} &=& (\lambda_{s};\forall s,\;\phi_{k};\forall k,\;\chi_{s})^T  \\
  {\bfg \mu} &=& (\phi_{k}^{\rm max}; \forall k,\;\phi_{k}^{\rm min};\forall k,\; \xi_{s}^{\rm max};\forall s,\;\xi_{s}^{\rm min};
\forall s )^T \\
  {\bfg \alpha} &=& (\alpha_{j}^{\rm D}; \forall j )^T \\
  {\bfg \varphi} &=& -(\varphi_{i}^{\rm G};\forall i\;, \varphi_{j}^{\rm D}; \forall j)^T.
\end{eqnarray}

\section{Generation and load data for the IEEE-118 test system}\label{IEEE-118data}

Table~\ref{busdata118} provides the location of generators and demands in the network, the maximum power and the corresponding cost for generating units, and the load and nominal load-shedding cost for demands.
\begin{table}[htb]
\renewcommand{\arraystretch}{1.0}
\renewcommand{\tabcolsep}{1.0pt}
{\scriptsize
\begin{center}
\caption{Generator and demand data for IEEE 118 case study}\label{busdata118}
\begin{tabular}{|c|c|c|c|c||c|c|c|c|c||c|c|c|c|c||c|c|c|c|c|}\hline
& \multicolumn{2}{c|}{Generators} & \multicolumn{2}{c|}{Demands} & & \multicolumn{2}{c|}{Generators} & \multicolumn{2}{c|}{Demands} & &\multicolumn{2}{c|}{Generators} & \multicolumn{2}{c|}{Demands}& & \multicolumn{2}{c|}{Generators} & \multicolumn{2}{c|}{Demands} \\ \hline
Bus & Pow. & Cost & Load & NLSC & Bus & Pow. & Cost & Load & NLSC & Bus & Pow. & Cost & Load & NLSC & Bus & Pow. & Cost & Load & NLSC\\\hline\hline
%& (MW) & (\euro/MWh) & (MW) & (\euro/MWh) & & (MW) & (\euro/MWh) & (MW) & (\euro/MWh)& (MW) & (\euro/MWh) & (MW) & (\euro/MWh)& (MW) & %(\euro/MWh) & (MW) & (\euro/MWh)\\ \hline
1	&	--	&	--	&	689	&	50	&	31	&	150	&	27	&	581	&	50	&	61	&	1000	&	14&	--	&	--		&	91	&	250	&	23&	--	&	--	 \\\hline	
2	&	--	&	--	&	270	&	50	&	32	&	500	&	18	&	797	&	50	&	62	&	500	&	18	&	1040	&	50	&	92	&	1500	&	14	&	878	&	 50	 \\\hline
3	&	--	&	--	&	527	&	50	&	33	&	--	&	--	&	311	&	50	&	63	&	--	&	--	&	--	&	--	&	93	&	--	&	--	&	162	&	50	 \\\hline
4	&	150	&	27	&	405	&	50	&	34	&	150	&	27	&	797	&	50	&	64	&	--	&	--	&	--	&	--	&	94	&	--	&	--	&	405	&	50	 \\\hline
5	&	--	&	--	&	--	&	--	&	35	&	--	&	--	&	446	&	50	&	65	&	2100	&	10&	--	&	--		&	95	&	--	&	--	&	567	&	 50	 \\\hline
6	&	150	&	27	&	702	&	50	&	36	&	500	&	18	&	419	&	50	&	66	&	2100	&	10	&	527	&	50	&	96	&	--	&	--	&	513	&	 50	 \\\hline
7	&	--	&	--	&	257	&	50	&	37	&	--	&	--	&	--	&	--	&	67	&	--	&	--	&	378	&	50	&	97	&	--	&	--	&	203	&	50	 \\\hline
8	&	150	&	27&	--	&	--		&	38	&	--	&	--	&	--	&	--	&	68	&	--	&	--	&	--	&	--	&	98	&	--	&	--	&	459	&	50	 \\\hline
9	&	--	&	--	&	--	&	--	&	39	&	--	&	--	&	365	&	50	&	69	&	1500	&	14&	--	&	--		&	99	&	1500	&	14&	--	&	 --	\\	 \hline
10	&	1500	&	14&	--	&	--		&	40	&	150	&	27	&	270	&	50	&	70	&	400	&	17	&	891	&	50	&	100	&	1500	&	14	&	500	&	 50	 \\\hline
11	&	--	&	--	&	945	&	50	&	41	&	--	&	--	&	500	&	50	&	71	&	--	&	--	&	--	&	--	&	101	&	--	&	--	&	297	&	50	 \\\hline
12	&	1500	&	14	&	635	&	50	&	42	&	150	&	27	&	500	&	50	&	72	&	150	&	27&	--	&	--		&	102	&	--	&	--	&	68	&	 50	 \\\hline
13	&	--	&	--	&	459	&	50	&	43	&	--	&	--	&	243	&	50	&	73	&	150	&	27&	--	&	--		&	103	&	100	&	38	&	311	&	50	 \\\hline
14	&	--	&	--	&	189	&	50	&	44	&	--	&	--	&	216	&	50	&	74	&	100	&	38	&	918	&	50	&	104	&	500	&	18	&	513	&	50	 \\\hline
15	&	150	&	27	&	1215	&	50	&	45	&	--	&	--	&	716	&	50	&	75	&	--	&	--	&	635	&	50	&	105	&	500	&	18	&	419	&	 50	 \\\hline
16	&	--	&	--	&	338	&	50	&	46	&	500	&	18	&	378	&	50	&	76	&	500	&	18	&	918	&	50	&	106	&	--	&	--	&	581	&	50	 \\\hline
17	&	--	&	--	&	149	&	50	&	47	&	--	&	--	&	459	&	50	&	77	&	500	&	18	&	824	&	50	&	107	&	100	&	38	&	378	&	50	 \\\hline
18	&	500	&	18	&	810	&	50	&	48	&	--	&	--	&	270	&	50	&	78	&	--	&	--	&	959	&	50	&	108	&	--	&	--	&	27	&	50	 \\\hline
19	&	150	&	27	&	608	&	50	&	49	&	1250	&	13	&	1175	&	50	&	79	&	--	&	--	&	527	&	50	&	109	&	--	&	--	&	108	&	 50	 \\\hline
20	&	--	&	--	&	243	&	50	&	50	&	--	&	--	&	230	&	50	&	80	&	1500	&	14	&	1755	&	50	&	110	&	250	&	23	&	527	&	 50	 \\\hline
21	&	--	&	--	&	189	&	50	&	51	&	--	&	--	&	230	&	50	&	81	&	--	&	--	&	--	&	--	&	111	&	500	&	18&	--	&	--	\\	 \hline
22	&	--	&	--	&	135	&	50	&	52	&	--	&	--	&	243	&	50	&	82	&	500	&	18	&	729	&	50	&	112	&	500	&	18	&	338	&	50	 \\\hline
23	&	--	&	--	&	95	&	50	&	53	&	--	&	--	&	311	&	50	&	83	&	--	&	--	&	270	&	50	&	113	&	500	&	18&	--	&	--	\\	 \hline
24	&	150	&	27&	--	&	--		&	54	&	1250	&	13	&	1526	&	50	&	84	&	--	&	--	&	149	&	50	&	114	&	--	&	--	&	108	&	 50	 \\\hline
25	&	1500	&	14&	--	&	--		&	55	&	500	&	18	&	851	&	50	&	85	&	150	&	27	&	324	&	50	&	115	&	--	&	--	&	297	&	 50	 \\\hline
26	&	1750	&	11&	--	&	--		&	56	&	500	&	18	&	1134	&	50	&	86	&	--	&	--	&	284	&	50	&	116	&	250	&	23&	--	&	 --	 \\\hline	
27	&	150	&	27	&	837	&	50	&	57	&	--	&	--	&	162	&	50	&	87	&	1500	&	11&	--	&	--		&	117	&	--	&	--	&	270	&	 50	 \\\hline
28	&	--	&	--	&	230	&	50	&	58	&	--	&	--	&	162	&	50	&	88	&	--	&	--	&	648	&	50	&	118	&	--	&	--	&	446	&	50	 \\\hline
29	&	--	&	--	&	324	&	50	&	59	&	1000	&	14	&	3739	&	50	&	89	&	1000	&	14&	--	&	--	& --	&	--	&	--	&	--	&	 --	\\	 \hline							
30	&	--	&	--	&	--	&	--	&	60	&	--	&	--	&	1053	&	50	&	90	&	100	&	38	&	1053	&	50 & --	&	--	&	--	&	--	&	 --	 \\\hline	
\multicolumn{20}{l}{Offer (MW), bid (MW), price (\euro/MWh), NLSC: Nominal Load-Shedding Cost}\\ 	\hline										
\end{tabular}
\end{center}
}
\end{table}

\bibliographystyle{elsarticle-num}      % basic style, author-year citations
%\bibliography{BiblioIngAll}

\end{document}